\documentclass[twoside,english,prl,twocolumn,showpacs,superscriptaddress,reprint]{revtex4-1}

\usepackage[T1]{fontenc}
\usepackage[latin9]{inputenc}
\usepackage{geometry}
\usepackage{amsmath}
\usepackage{amssymb}
\usepackage[pdftex]{graphicx}
\usepackage{times}

\begin{document}

\title{Wettability stabilizes fluid invasion into porous media via nonlocal, cooperative
pore filling}

\author{Ran Holtzman}
\email{holtzman.ran@mail.huji.ac.il}
\address{Department of Soil and Water Sciences, The Hebrew University of Jerusalem,
Israel}
\author{Enrico Segre}
\address{Physics Services, Weizmann Institute of Science, Israel}

\date{\today}

\begin{abstract}
We study the impact of the wetting properties on the immiscible displacement
of a viscous fluid in disordered porous media. We present a novel pore-scale model that captures wettability and dynamic effects, including the spatiotemporal nonlocality associated with interface readjustments. Our simulations show that increasing the wettability of the invading fluid (the contact angle) promotes cooperative pore filling that stabilizes the invasion, and that this effect is suppressed as the flow rate increases, due to viscous instabilities. We use scaling analysis to derive two dimensionless numbers that predict the mode of displacement. By elucidating the underlying mechanisms, we explain classical yet intriguing experimental observations. These insights could be used to improve technologies such as hydraulic fracturing,
CO$_{2}$ geo-sequestration, and microfluidics. 
\end{abstract}


\maketitle


Fluid-fluid displacement in porous media is important in natural and
industrial processes at various scales, from enhanced energy recovery,
CO$_{2}$ geo-sequestration, groundwater contamination and soil wetting
and drying, to dyeing of paper or textiles and microfluidics. 
Fluid displacement is governed by the interplay between quenched disorder, short-range cooperative effects and long-range pressure screening, which depends on a large number of parameters, including the wettability---the relative affinity of the fluids to the solid. Consequently, the displacement patterns can range from a stable, compact front to highly ramified with preferential flow paths (fingers)~\cite{Alava2004}.
Fluid invasion is a member of a broad class of problems characterized by competitive domain growth and nonlinear interface dynamics, including magnetic domains, biological films and flame front propagation~\cite{Pelce2004}.
The interface evolution in these systems is often modeled as a competition between the energy associated with the interaction between phases and constraints arising from disorder; the relative importance of the two can be tuned by properties such as wettability in fluid displacement or local random interaction fields in magnetic domains~\cite{Ji1991}.
Understanding the impact of wettability on fluid invasion---the topic of this Letter---is therefore relevant to a wide range of phenomena of scientific and technological importance.

Immiscible displacement can be classified according to the wettability into drainage or nonwetting invasion, where the displaced fluid preferentially wets the solid (contact angle $\theta < 90 \textdegree$, measured through the defending fluid), or imbibition of a wetting fluid ($\theta>90\textdegree$). Intensive research has provided basic understanding of drainage, identifying different invasion behaviors and explaining their dependence
on the flow velocity, fluid viscosities, interfacial tension, and
the degree of pore-scale disorder (\cite{lenormandtouboul88,lenormand90-liquids,toussaintlovoll05,holtzmanjuanes10-fingfrac} and the references therein).
Increasing $\theta$ was found to stabilize the displacement and reduce
trapping in forced and gravity-driven drainage experiments~\cite{Cottin2011, Shahidzadeh-Bonn2004}.

In contrast, relatively few works have studied imbibition,
mostly for the stable case of a more viscous invading fluid~\cite{lenormandtouboul88, lenormand90-liquids, He1992}. 
For unstable viscosity ratios, experiments showed a marked difference between viscous fingering in drainage and more stable patterns with thicker fingers in imbibition~\cite{stokesweitz86}. Stabilization was also captured in simulations which introduced viscous effects stochastically
\cite{Rangel2009} and in lattice Boltzmann simulations~\cite{Liu2013}. 
Quasi-static simulations (neglecting dynamic effects) illustrated that increasing $\theta$ enhanced the occurrence of a nonlocal, cooperative pore filling mechanism, resulting in a compact pattern~\cite{Cieplak1988,*Cieplak1990}.
These intriguing results were only recently explored systematically by experiments in which the wettability was altered while keeping the same fluid pair~\cite{Trojer_PRAP2015}. The authors demonstrated that increasing $\theta$ stabilized the displacement, leading to a compact front in slow imbibition despite the high, unfavorable viscosity ratio~\cite{Trojer_PRAP2015}. 
Many of these important observations remain unexplained, primarily because of nonlocal pore filling dynamics, that is inaccessible experimentally and not well-characterized by existing models~\cite{Meakin2009, Trojer_PRAP2015}. 
In this Letter, we present a novel pore-scale model that exposes the competing effects of wettability and flow rate, thereby explaining the aforementioned observations.

We develop a two-dimensional, discrete model of immiscible displacement in a random medium with fluids of
arbitrary viscosities and contact angle.
Our model is briefly described below, and in further details as Supplemental Material~\cite{SuppMat}.
A mechanistic description of the displacement dynamics with both capillary and viscous forces is obtained by combining two modeling approaches: (a) grain-based
\cite{Cieplak1988,Cieplak1990,Motealleh2010}, resolving meniscus
stability from pore geometry; and (b) pore-based~\cite{Blunt1998,holtzmanjuanes10-fingfrac},
resolving fluid pressures and fluxes from the pore topology and geometry. 
Through consideration of viscous dissipation, our model captures the nonlocal nature of interface dynamics: the effect of local pore invasion on the interface configuration elsewhere, the disparate timescales of pore filling and bulk flow~\cite{Armstrong2013, Berg2013}, and the associated mechanisms of pressure screening~\cite{niemeyerpietronero84,lovollmeheust04} and interface readjustments~\cite{ furubergmaloy96, Armstrong2013}. These mechanisms are crucial even in slowly-driven systems, limiting the volume invaded in a single event (avalanche)~\cite{furubergmaloy96, Armstrong2013} and increasing trapping~\cite{Joekar-Niasar_CREST_2012}, and, for the conditions considered here---high disorder and porosity, small throat to pore size ratio, and large viscosity ratio---have a stronger impact on the displacement than other mechanisms such as film flow, snapoff, and contact angle variations~\cite{lenormand90-liquids, Alava2004,Joekar-Niasar_CREST_2012}. Furthermore, modeling contact line and contact angle dynamics is strongly debated, and requires consideration of details down to the molecular level~\cite{Alava2004,Meakin2009}. 
Considering the relative impact of these mechanisms, as well as the complexity and ambiguity involved in their implementation, we emphasize viscosity-related mechanisms and exclude liquid films and contact line/angle dynamics (see~\cite{SuppMat} for elaborated discussion). 
Consequently, our model provides the coupled effects of wettability and dynamics in large, disordered domains, improving upon existing models which either ignore dynamics and/or wettability effects, or are limited by computational cost to small domains~\cite{Meakin2009, Joekar-Niasar_CREST_2012}.

\begin{figure}
\centering
\includegraphics[width=1\columnwidth]{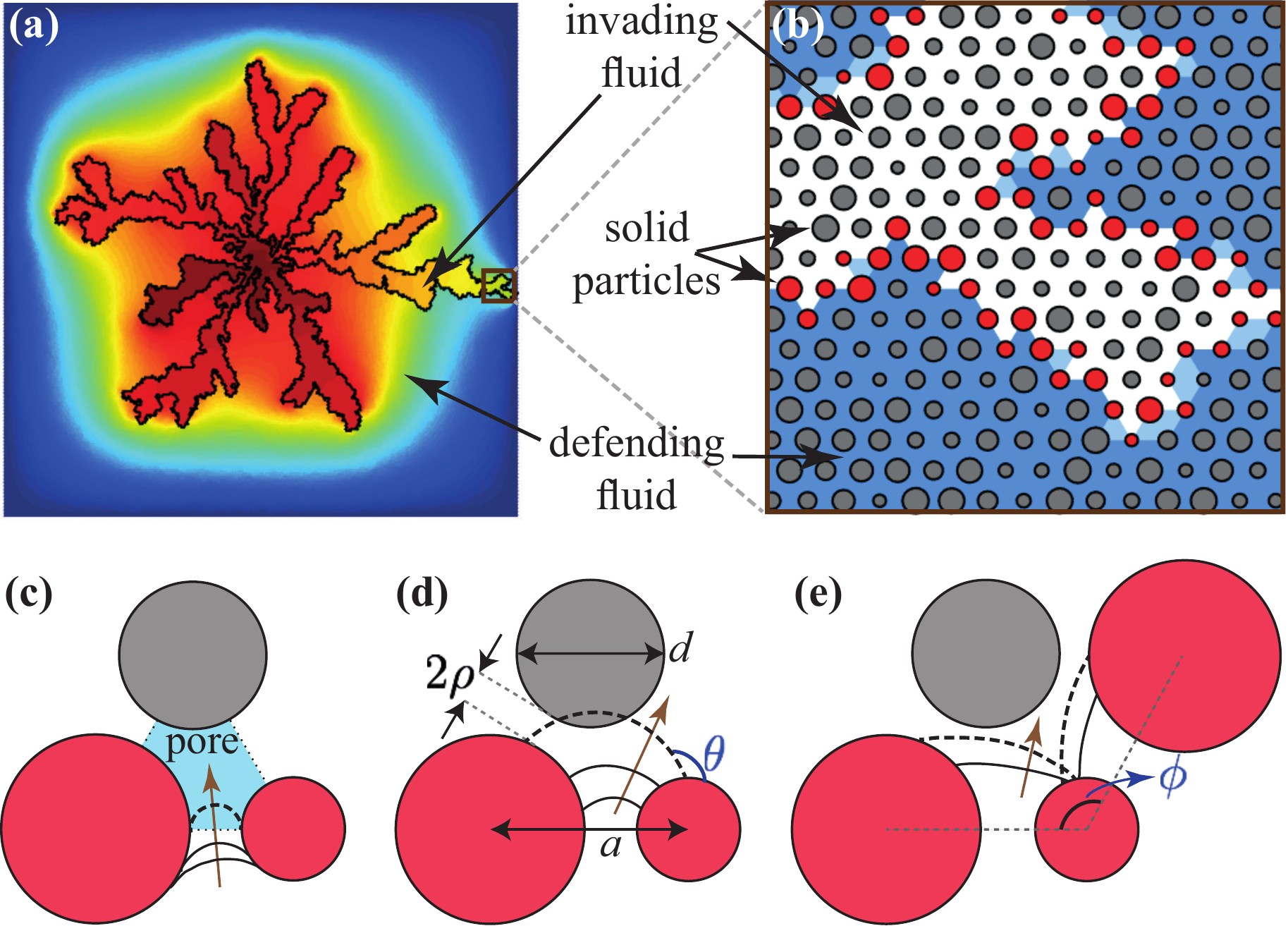}
\caption{(color online). Model schematic. (a) We simulate radial displacement, tracking the fluid-fluid interface (black line) and fluid pressures (increasing from blue to red).
(b) Zoom in showing the lattice of particles and pores. The interface is represented as a sequence of circular menisci, touching particles at contact angle $\theta$, with curvature set by the local capillary pressures. Menisci can be destabilized by: (c) {burst}; (d) {touch}; or (e) overlap. Brown arrows indicate direction of advancement, destabilized arc in dash.
\label{fig:model_schematics}}
\end{figure}

We construct a disordered
medium by placing cylindrical solid particles on a triangular
lattice (spacing $a$), selecting the particle diameters $d$
from an assigned distribution; here uniform, $d\in[1-\lambda,1+\lambda]\bar{d}$,
where $\lambda\in(0,1)$ is the degree of disorder, and $\bar{d} \lesssim a$
is the mean diameter. The triangular cell delimited by a particle triplet
defines a pore of volume $V$, connected to three neighbors
via throats of width $2\rho\lesssim a$ [Fig.~\ref{fig:model_schematics}]. 
The fluid-fluid interface is represented by a sequence of circular arcs (menisci); each arc intersects a
pair of particles at the prescribed contact angle $\theta$~\footnote{We define $\theta$ as an effective contact angle, representing here the advancing angle, which could be affected by flow rates and altered if a thin film travels ahead of the imbibing front.},
with a radius of curvature $R\sim\gamma/\Delta p$
related to the capillary pressure $\Delta p$ via the Young-Laplace law, where $\gamma$ is the interfacial tension.

We consider three types of capillary instabilities~\cite{Cieplak1988,*Cieplak1990}: (1)
Haines jump or \emph{burst}, when the curvature exceeds a threshold;
(2) \emph{touch}, when a meniscus intersects a third particle; and
(3) \emph{overlap} of adjacent menisci, destabilizing each other [Fig.~\ref{fig:model_schematics}(c--e)]. Overlap (termed ``Melrose event''
in~\cite{Motealleh2010}) is a nonlocal, cooperative mechanism affected
by the menisci in multiple pores, smoothing the interface~\cite{Cieplak1988,*Cieplak1990}. 
%
%
%

Meniscus instability causes its incipient advancement. Both stability and advancement rate depend upon the pressure difference across each meniscus. 
Pore pressures and filling rates are provided by the fluids' viscous resistance, evaluated from the flow throughout the network of contiguous pores occupied by same fluid and through throats with unstable, advancing menisci. Flow is resolved via conservation of fluid mass in each pore, $\sum_j q_j=0$ (summing over all neighboring pores $j$). 
Assuming Stokes flow, $q = C \nabla p $ provides the interpore flow rate, where $C \sim\rho^{4} / \mu\mathrm{_{eff}} $ is the conductance and $\nabla p = (p_j-p) / \rho$. 
An effective viscosity, $\mu\mathrm{_{eff}}=\left(\mu_{i}-\mu_{d}\right)\varPhi+\mu_{d}$ allows using $q$ to evaluate both flow of a single fluid between two pores and filling rate~\cite{lenormandtouboul88}. 
Here $\mu_{d}$ and $\mu_{i}$ are the defending and invading fluid
viscosities. The filling status of the invaded
pore, $0\le\varPhi\le1$, is updated according to the inflow from throats with unstable menisci $q^\mathrm{inv} = \sum_u q_u$ at each time step $t$,
$\varPhi(t+\Delta t)=\varPhi(t)+q^\mathrm{inv}(t)\Delta t/V$. 
Front readjustments are incorporated by considering partially-filled pores, which can re-empty upon reversal of
meniscus advancement direction. 
When pore invasion is completed ($\varPhi=1$), the interface configuration is updated~\cite{SuppMat}. 
%
The above provides a simple description of the invasion dynamics without explicit geometrical calculations of changes in fluid volume from changes in menisci curvature.
We enforce a constant injection rate from a radial region of several pores (inlet), stopping the simulations when a boundary (outlet) pore is invaded.



\begin{figure}
\centering
\includegraphics[width=1\columnwidth]{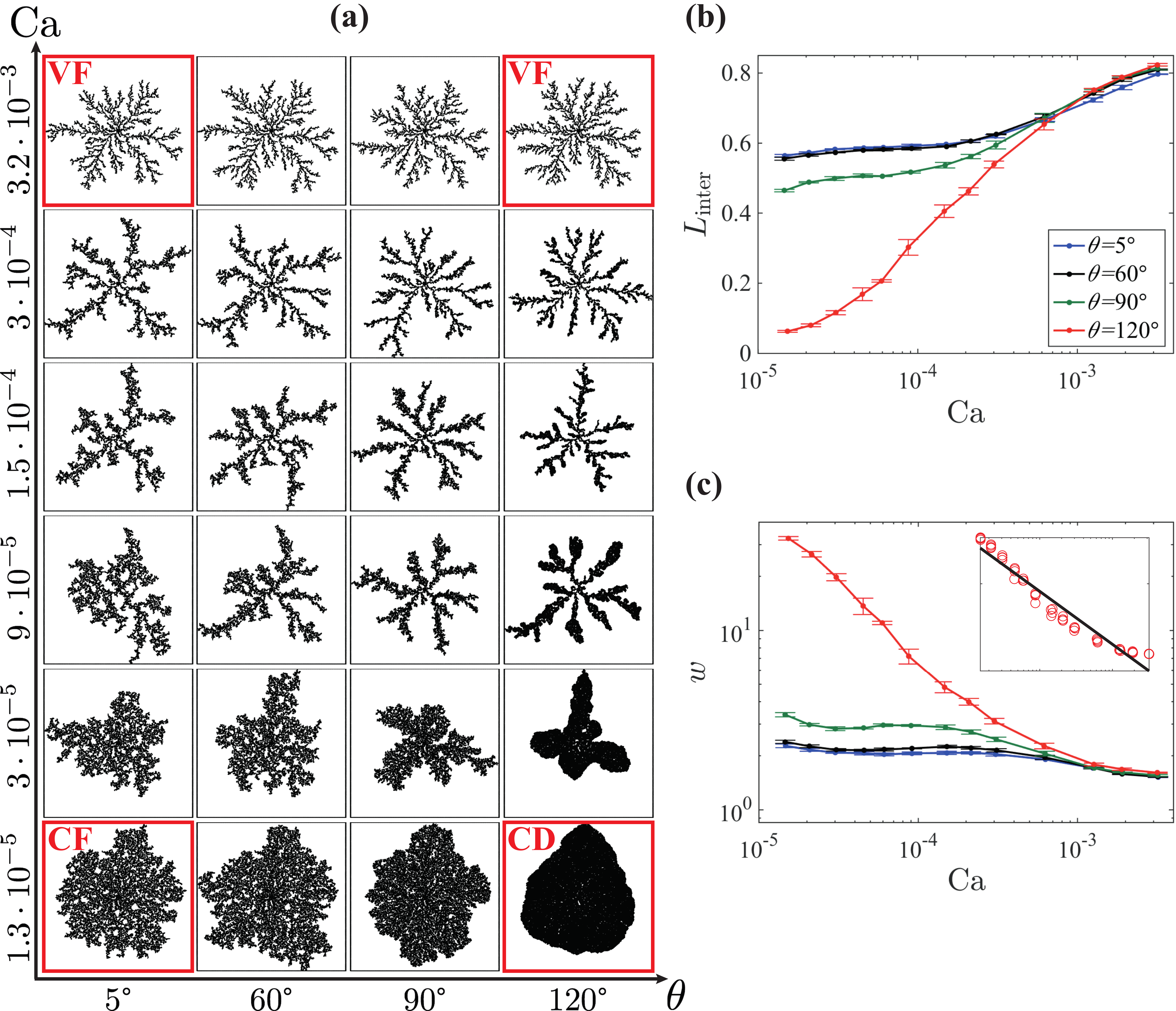}
\caption{(color online). (a) Simulated invasion patterns, characterized by: (b) interface length,
$L\mathrm{_{inter}}$, and (c) finger width, $w$. Rapid
injection (high $\mathrm{Ca}$) leads to viscous fingering (VF)
with irregular interfaces ($L\mathrm{_{inter}}\approx1$) and thin fingers ($w\approx1$).
As $\mathrm{Ca}$ is decreased, the patterns transition towards capillary
fingering (CF) with multiple trapped clusters and relatively long
interfaces in drainage (low $\theta$), or compact displacement (CD,
$L\mathrm{_{inter}}\ll1$, $w\gg1$) in imbibition (high $\theta$). 
The inset of (c) shows a fit of $w\sim\mathrm{Ca^{-\nu}}$ for $\theta = 120 \textdegree$ providing $\nu \approx 0.6$.
Error bars show the standard deviation among four realizations.
\label{fig:patterns}}
\end{figure}

Our simulations exhibit the experimentally-observed invasion regimes
\cite{Trojer_PRAP2015}: viscous fingering in rapid injection
irrespective of the wettability, capillary fingering in slow drainage, and stable, compact displacement
in slow imbibition [Fig.~\ref{fig:patterns}(a)].
For a fixed fluid pair (constant $\mu_{d} / \gamma$), the dimensionless flow rate is provided by the capillary number, $\mathrm{Ca}=\mu_{d}v/\gamma$, computed
from the velocity $v=(V{}_{tot}/t{}_{tot})/A{}_{out}$, where $V{}_{tot}$
is the volume drained during the simulation time $t_{tot}$, through
the outlet cross-sectional area $A_{out}$. 
To simulate the injection of air into water-glycerol saturated beads in~\cite{Trojer_PRAP2015}, we used the following parameters:
$\gamma=67\cdotp10^{-3}$ N/m, $\mu_{i}=1.8\cdotp10^{-5}$ Pa·s, $\mu\mathrm{_{d}}=5\cdotp10^{-3}$
Pa·s, $a=500\:\mu m$, $\bar{d}=0.54a$, system size $L=260a$
(260$\times$300 particles), and $\lambda=0.81$
(providing the wide variation of aperture sizes in random bead packs~\cite{Cieplak1988,*Cieplak1990}).

We characterize the patterns
quantitatively via the length of the fluid-fluid interfaces (including
trapped regions) normalized by the invaded area, $L\mathrm{_{inter}}$~\footnote{The length of the fluid-fluid interfaces is computed from the number of interfacial pores divided by the number of invaded pores; $L\mathrm{_{inter}} \rightarrow 1$ for thin fingers, and $L\mathrm{_{inter}} \rightarrow L^{-1}$ for a compact front ($L$ being the system size).}, and the mean finger width $w$ (in lattice
units $a$~\footnote{We evaluate the finger width by two methods: (a) skeletonizing the invasion pattern with a Voronoi algorithm, and measuring the distance of the medial line from the interface [e.g. see R. L. Ogniewicz and O. Kubler, Pattern Recognition \textbf{28}, 343 (1995)]; and (b) measuring the contiguous invaded length along parallel cuts~\cite{Cieplak1988,*Cieplak1990}. The two methods provided similar values.}).
Viscous fingering is characterized
by thin fingers of a single pore width, $w\approx1$, and long, highly irregular interfaces,
$L\mathrm{_{inter}}\approx1$, whereas compact displacement provides a smooth, rounded
front, $L\mathrm{_{inter}}\ll1$, with a diverging finger width, $w\gg1$. In capillary fingering, trapping provides 
long, fractal interfaces, which a patchy, thick pattern composed of multiple thinner, contiguous fingers [Fig.
\ref{fig:patterns}(b--c)].
The robustness of our characterization 
is demonstrated by the consistency among four realizations (same particle size distribution). 
Our simulations capture the decrease in finger width with imbibition rate, providing $w\sim\mathrm{Ca^{-\nu}}$ with $\nu \approx 0.6$ [$\theta = 120 \textdegree$, see inset of Fig.~\ref{fig:patterns}(c)]. While the small difference from $\nu=0.51$ in~\cite{stokesweitz86} can be explained by the use of different fluids (and $\theta$), we note that saturation of $w \rightarrow 1$ at high $\mathrm{Ca}$ exacerbates the quality of fit. 
We also find that sweep efficiency decreased sharply between compact displacement, capillary and viscous fingering, however non-monotonically~\cite{Liu2013}.

\begin{figure}
\centering
\includegraphics[width=.7\columnwidth]{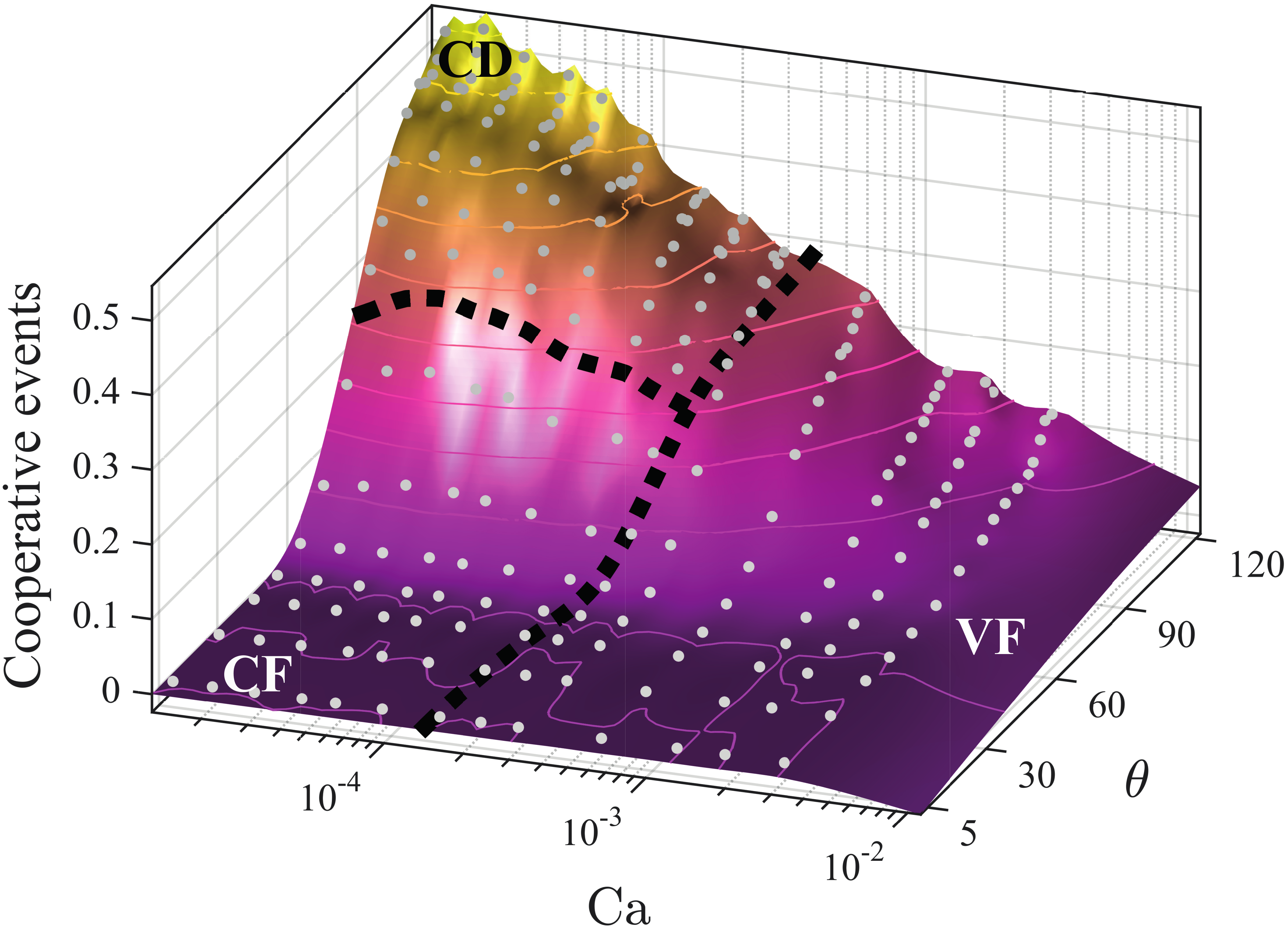}
\caption{(color online). Occurrence of nonlocal, cooperative pore
filling (number of overlaps out of all instability events)
from 208 simulations (gray
dots). Increasing $\theta$ enhances overlaps, manifested macroscopically by a more stable displacement. Dashed lines mark the theoretical phase boundaries predicted by scaling (see text).
\label{fig:mechanisms}}
\end{figure}

The crossover between the invasion regimes depends on the interplay
between three mechanisms: (i) continuous growth of thin
fingers; (ii) intermittent interface advancement at different locations, trapping the defending fluid behind; and (iii) simultaneous advancement of large parts
of the interface, keeping it smooth. 
Finger growth in (i) is driven by destabilization of the entire interface at high $\mathrm{Ca}$, where high defending fluid pressure in the ``gulfs'' between fingers allows only the finger tips to advance.
This screening effect [demonstrated by the pressure halo in Fig.~\ref{fig:model_schematics}(a)], where Laplacian-driven growth dominates over heterogeneity~\cite{niemeyerpietronero84,lovollmeheust04}, promotes viscous fingering
(see Videos 1a and 1b in~\cite{SuppMat}). 
In (ii), disorder in entry pressures leads to capillary fingering at low $\mathrm{Ca}$ and $\theta$ (Video
2 in~\cite{SuppMat}). At low $\mathrm{Ca}$ and high $\theta$, the dominance of overlaps (Fig.~\ref{fig:mechanisms}) enhances mechanism (iii), where invasion in one location destabilizes the interface in adjacent pores (Video 3 in~\cite{SuppMat}, experimentally observed in~\cite{Trojer_PRAP2015}), resulting in compact displacement.
We emphasize that although bursts are the dominant invasion mechanism at low $\theta$ irrespective of $\mathrm{Ca}$ (Fig.~\ref{fig:mechanisms}), the change in driving mechanism (from i to ii as $\mathrm{Ca}$ is decreased) leads to different patterns.

We rationalize the invasion behavior by evaluating the magnitude of the forces driving mechanisms (i)--(iii). We predict the transition between viscous fingering and capillary
fingering/compact displacement through a capillary number modified to account for the contact angle, $N\mathrm{_{Ca}}=\delta p_{\bot} / \delta p_{\Vert}$. Here $\delta p_{\bot}$ is the pressure drop driving growth of individual fingers
perpendicular to the interface (along the direction of the externally-applied pressure drop), and $\delta p_{\Vert}$ is the capillary pressure promoting lateral growth of the interface. We evaluate $\delta p_{\bot}$
from the pressure drop in the viscous defending fluid over a characteristic
length $L_{\bot}$, $\delta p_{\bot}\sim\nabla p_{\bot}L_{\bot}$, where $\nabla p_{\bot}\sim\mu_{d}v/k$ with permeability $k \sim a^2$ and $L_{\bot} \sim a$. 
We use the critical burst curvature $R_{c}$~(Eq.~S2 in \cite{SuppMat}) to evaluate the capillary force, $\delta p_{\Vert}\sim\gamma/R_{c}$, providing 
\begin{equation}
{N\mathrm{_{Ca}}}=\mathrm{Ca} \left( \sqrt{1-\tilde{l}{}^{2}\sin{}^{2}\theta} - \tilde{l}\cos\theta \right),\label{eq:NCa}
\end{equation}
where $\tilde{l}=\bar{d}/a$ is the dimensionless microscopic characteristic length.

For slow injection, we explain the transition between capillary fingering and compact displacement 
via the ``Cooperative number'' $N_\mathrm{coop}$, a
dimensionless parameter evaluating the likelihood for pore filling
by overlaps, 
\begin{equation}
N_\mathrm{coop} = \cos\frac{\phi}{2} - \tilde{l}\sin{}^{2}\theta + \cos\theta\sqrt{1-\tilde{l}{}^{2}\sin{}^{2}\theta},\label{eq:Ncoop}
\end{equation}
where $N_\mathrm{coop}=0$ is the geometrical condition for two
arcs to overlap \emph{exactly} at their threshold (burst) curvature, such that $N_\mathrm{coop}>0$ implies overlap \emph{preceding} burst~\cite{SuppMat}.
Here $\phi$ is the local front shape, defined by the angle between two adjacent menisci [Fig. \ref{fig:model_schematics}(e)]. Since the macroscopic pattern is a consequence of numerous invasion
events occurring at $\phi$ which vary in time
and space, $N_\mathrm{coop}$ represents the instability statistics of the entire sample and simulation time:
a larger $N_\mathrm{coop}$ value implies a higher fraction of overlaps; said differently, for a given
$N_\mathrm{coop}$ value not all pores will be invaded by the same
instability (Fig.~\ref{fig:mechanisms}). Here we compute $N_\mathrm{coop}$ using $\phi=120 {\textdegree}$, which we found to be most representative for our system~\cite{SuppMat}.

\begin{figure}
\centering
\includegraphics[width=1\columnwidth]{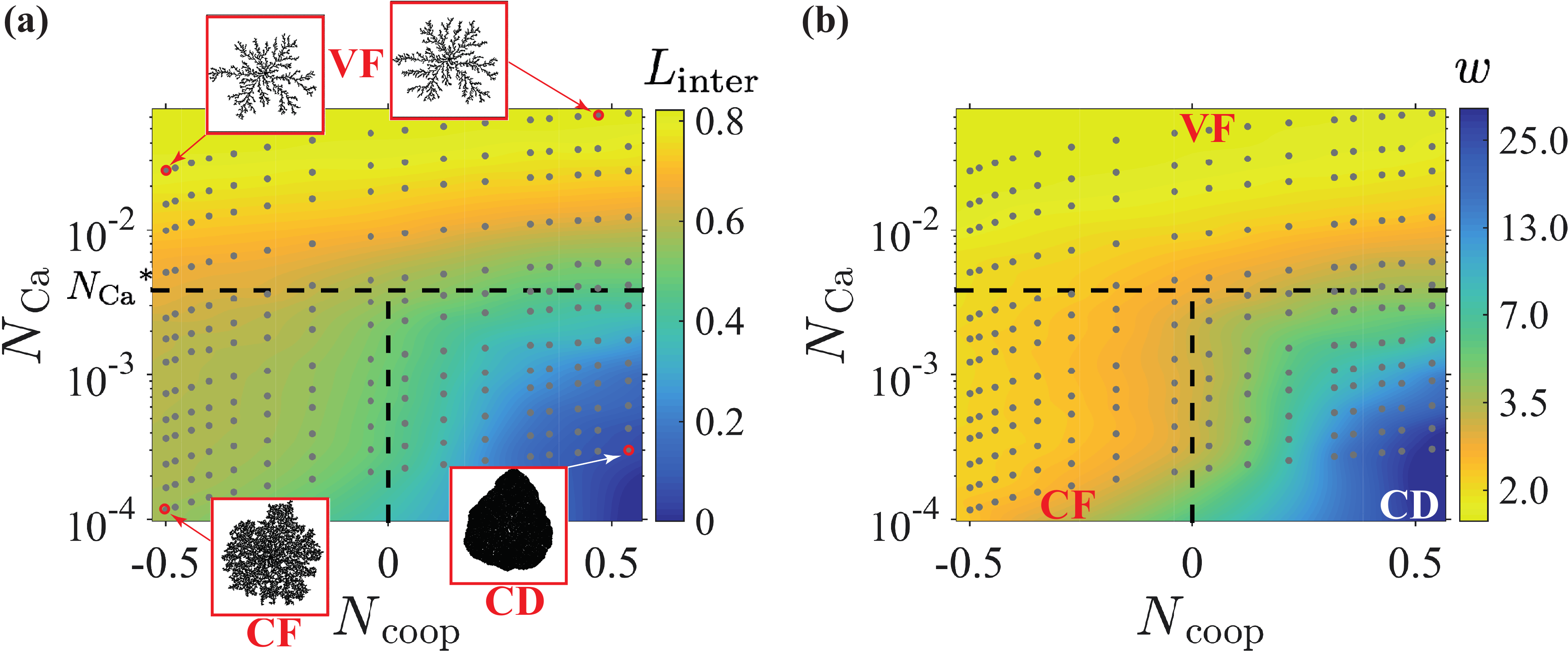}
\caption{(color online). Phase diagrams of immiscible displacement: (a) interface length, $L\mathrm{_{inter}}$ and (b) finger width, $w$.
At high flow rates, $N_\mathrm{Ca} \gg {N_\mathrm{Ca}}^*$
predicts viscous fingering (VF) with long, fractal interfaces and thin fingers ($L\mathrm{_{inter}}$$\approx$$1$, $w$$\approx$$1$). At
low rates, $N_\mathrm{Ca} \ll {N_\mathrm{Ca}}^*$, invasion is controlled by the wettability: for drainage, $N_\mathrm{coop}<0$
implies capillary fingering (CF), whereas for imbibition $N_\mathrm{coop}>0$ indicates compact
displacement (CD, $L\mathrm{_{inter}}$$\ll$$1$, $w$$\gg$$1$) due to smoothing by cooperative pore filling.
Dashes show phase boundaries from scaling analysis, $N_\mathrm{Ca} = {N_\mathrm{Ca}}^* \approx 4\cdotp10^{-3}$ and $N_\mathrm{coop} = 0$. Dots mark data from 208 simulations at various $\mathrm{Ca}$ and $\theta$.
\label{fig:scaling_phase_diagram}}
\end{figure}

Our scaling analysis predicts the mode of invasion. For rapid injection, $N_\mathrm{Ca} \gg {N_\mathrm{Ca}}^*$
implies dominance of viscous forces leading to viscous fingering. Here, the critical value scales as ${N_\mathrm{Ca}}^* \sim (L/a)^{-1} \approx 4\cdotp10^{-3}$, suggesting a dependence on the macroscopic characteristic length---the system size~\cite{toussaintlovoll05,holtzmanjuanes10-fingfrac}.  
For slow injection, $N_\mathrm{Ca} \ll {N_\mathrm{Ca}}^*$, capillary forces govern and
invasion becomes strongly dependent on the wettability: for nonwetting invasion,
$N_\mathrm{coop}<0$ predicts capillary fingering caused by disorder
in capillary (burst) thresholds, whereas for wetting invasion $N_\mathrm{coop}>0$
implies cooperative motion of large parts of the interface (overlaps)
and a compact pattern,
in agreement with our simulations (Fig.~\ref{fig:scaling_phase_diagram}) and experiments~\cite{Trojer_PRAP2015}.
%

The displacement depends on the underlying medium geometry, including disorder, mean particle size and porosity, in a nontrivial manner, as it affects, together with wettability and flow rate, the portion of the pore space sampled by invasion.
%
For example, more pores would be invaded as disorder and $\mathrm{Ca}$ are decreased, while decreasing $\mathrm{Ca}$ and increasing $\theta$ restricts invasion to smaller pores. 
For the current geometry with high porosity ($\sim$0.67) and disorder ($\lambda=0.81$), $N_\mathrm{Ca}$ is relatively insensitive to $\theta$, increasing by a factor of $\sim$2.5 from $5 {\textdegree}$ to $120 {\textdegree}$. 
According to Eqs.~(\ref{eq:NCa}--\ref{eq:Ncoop}), the sensitivity of $N_\mathrm{Ca}$ to $\theta$ increases and the threshold angle (corresponding to $N_\mathrm{coop}=0$, here $\theta = 87 {\textdegree}$) decreases with particle size $\tilde{l}$.

In this Letter, we have studied the unstable case of high disorder and viscosity ratio. 
Noteworthy perspectives, which we intend to study with our model, include the impact of the medium geometric properties, viscosity ratio, gravity, and matrix deformations. 
Our preliminary simulations suggest that decreasing the disorder stabilizes the displacement, in agreement with~\cite{Cieplak1988, Cieplak1990, lenormand90-liquids}. 
Increased stability is also expected by decreasing the viscosity ratio~\cite{lenormandtouboul88,lenormand90-liquids,Liu2013} or introducing gravity~\cite{Shahidzadeh-Bonn2004}. 
Particularly interesting is the coupling with fracturing and particle rearrangements, which
significantly affects nonwetting invasion into granular media~\cite{holtzmanjuanes10-fingfrac,Sandnes2011,Holtzman_prl2012}.


In conclusion, we elucidate the combined impact of wettability
and dynamics on immiscible displacement in disordered media.
Our novel model provides the spatiotemporal nonlocal effects of interface dynamics, which are crucial even for slow flows due to the intrinsic timescale of interfacial jumps which can be orders of magnitude smaller than of the bulk flow~\cite{Armstrong2013}, thereby explaining classical yet unresolved observations. 
We show that increasing the wettability of the invading fluid promotes cooperative pore filling that stabilizes the invasion, and that this effect weakens as flow rate increases and viscous instabilities become dominant. 
Our analysis quantifies the competition between mechanisms governing the displacement stability, insight that could be exploited in technologies such as microfluidics, hydraulic fracturing and oil recovery~\cite{Wu2010,Shahidzadeh-Bonn2004,Courbin2007}.
%
Furthermore, our approach---a set of local rules providing a minimal description of the microscopic physics in a sufficiently-large domain to capture the emergent macroscopic behavior---could provide a new modeling paradigm for other problems of front propagation in disordered media, in which competition between local disorder, short-range cooperativity and global screening play a role, such as active media and spin glasses~\cite{Pelce2004}.

\begin{acknowledgments}
R.H. gratefully acknowledges financial support by the Israeli Science
Foundation (ISF-867/13), United States-Israel Binational Science
Foundation (BSF-2012140) and Israel Ministry of Agriculture
and Rural Development (821-0137-13). The authors thank L. Goehring for helpful comments. 
\end{acknowledgments}

\bibliographystyle{apsrev4-1}
\bibliography{Wettability_Letter}

\end{document}